\documentstyle[psfig,draft]{mn}
\newcommand{\nuul}{\nu_{ul}}

\newcommand{\zta}{z_{\rm ta}}
\newcommand{\beq}{\begin{equation}}
\newcommand{\enq}{\end{equation}}

\begin{document}

%
\title{Lithium hydride in the early Universe and in protogalactic clouds}
\author [E. Bougleux and D. Galli]
        {Elena Bougleux$^1$ and Daniele Galli$^2$\\
      $^1$Dipartimento di Astronomia, Universit\`a di Firenze, 
      Largo E. Fermi 5, I-50125 Firenze, Italy \\
      $^2$ Osservatorio Astrofisico di Arcetri, Largo E. Fermi 5, 
      I-50125 Firenze, Italy}

\date{Accepted .................
      Received .................
      in original form ................. }

 
\maketitle

\begin{abstract}

We examine the processes of formation and destruction of LiH molecules
in the primordial gas at temperatures $T\leq 5000$~K in the framework
of standard Friedmann cosmological models and we compute the optical
depth of the Universe due to elastic Thomson scattering of cosmic
background photons on LiH.  With the help of a simple model of
evolution of a spherical density perturbation we follow the linear
growth and the subsequent collapse of protogalactic clouds of various
masses, evaluating consistently their physical characteristics and
chemical composition.  We determine the expected level of anisotropy of
the cosmic background radiation due to the presence of LiH molecules in
high-redshift protogalactic clouds.  We conclude that LiH spectral
features generated in the primordial gas are hardly detectable with
current techniques and instrumentation.

\end{abstract}

 
\section{Introduction}

It has been suggested that a finite amount of molecules such as H$_2$,
H$_2^+$, H$_3^+$, HD, HD$^+$, H$_2$D$^+$, HeH$^+$, LiH and LiH$^+$ in
the primordial gas can induce in the cosmic background radiation (CBR)
both {\em spectral distortions} and {\em spatial anisotropies} (see
Dubrovich~1994 and Maoli~1994 for recent reviews).  For example, if the
CBR spectrum had some intrinsic distortion in the Wien region
originated at high redshift ($z\ga 100$) by whatsoever process of early
energy release, then molecules act as heat pumps, absorbing energy in
their rotovibrational levels and re-emitting this energy at lower
frequencies via rotational transitions (Dubrovich~1977, 1983). Clearly,
such a process can occur only if the CBR spectrum deviates from a pure
Planck profile.  Dubrovich \& Lipovka~(1995) have performed detailed
calculations of this effect for H$_2$D$^+$ molecules: the distortions
appear like narrow absorption and emission features in the CBR
spectrum.

Spatial anisotropies in the CBR can be produced by Thomson scattering
of photons on molecules (or electrons) located in protoclouds moving
with a peculiar radial velocity component $v_{\rm pec}$ (Sunyaev \&
Zel'dovich~1972, 1980; Dubrovich~1977, 1983; Maoli~1994; Maoli,
Melchiorri \& Tosti~1994).  If the abundance of molecules in the
primordial gas is sufficiently high, the cumulative effect of elastic
Thomson scattering can even create a ``curtain'' that blurs primordial
CBR anisotropies (Dubrovich~1993). The level of secondary anisotropies 
in the CBR temperature at a frequency $\nu$ is simply
\beq
\label{sz}
\frac{\Delta T}{T}=-\frac{v_{\rm pec}}{c}(1-e^{-\tau_\nu}),
\enq
where $\tau_\nu$ is the optical depth at frequency $\nu$ through the
cloud.  The advantage offered by molecules is that they possess much
larger photon scattering cross sections (by more than 10 orders of
magnitudes) than free electrons.  The derivation of the resulting
$\Delta T/T$ requires the solution of the equation of radiative
transfer in the expanding Universe, described in detail in Appendix A.

In addition, the presence of molecules in protogalactic clouds may
induce small fluctuations in the CBR temperature when the population of
rotovibrational levels deviates from the equilibrium (Boltzmann)
distribution corresponding to the actual radiation temperature $T_{\rm
r}$. Since the temperature of the gas $T_{\rm g}$ and the radiation
temperature are different for $z\la 10^3$, the populations of molecular
levels may differ from a Boltzmann distribution with temperature
$T_{\rm r}$ in clouds of sufficiently high density where collisional
excitation processes become significant.  The level of of anisotropy
produced in this case is proportional to the difference between $T_{\rm
r}$ and the excitation temperature of the molecular transition.

The efficiency of the coupling between CBR photons and the primordial
gas clearly depends on the chemical composition of the gas, and
therefore a description of the chemistry of the early Universe is 
necessary. A large number of papers has been devoted to this problem
(e.g. Dalgarno \& Lepp~1987; Palla~1988; Puy et al.~1993; Palla, Galli
\& Silk~1995).

\begin{figure}
\centerline{\psfig{figure=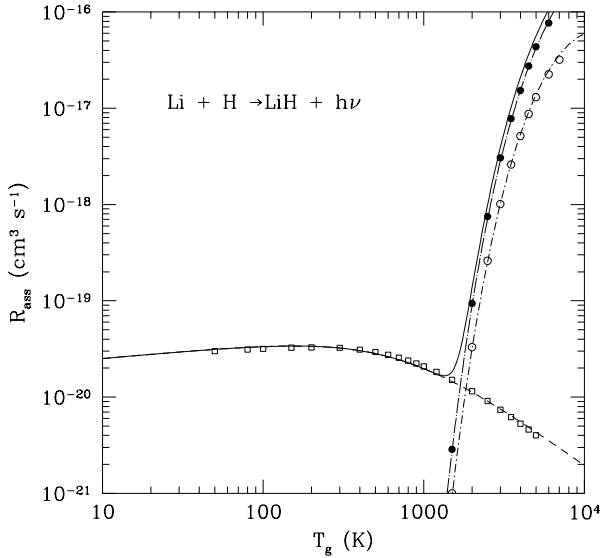,width=8.4cm}}
\caption[]{
The radiative association rate $R_{\rm ass}$ of LiH as function of the
gas temperature $T_{\rm g}$. The {\em dashed} line shows our fit to the
values computed by Dalgarno et al.~(1996) and Gianturco \& Gori
Giorgi~(1996a) (squares); the {\em dot long dashed} and {\em dot short
dashed} line our fit to the numerical results by Gianturco \& Gori
Giorgi~(1996b) (solid and empty circles) for the $B^1\Pi\rightarrow
X^1\Sigma^+$ and $A^1\Sigma^+\rightarrow X^1\Sigma^+$ processes
respectively. The relative populations of the $^2S$ and $^2P$ levels of
Li have been assumed to be at thermal equilibrium with temperature
$T_{\rm r}=T_{\rm g}$. The {\em solid} line shows the total rate.  
} 
\end{figure}

Of all primordial molecules, lithium hydride (LiH) is of considerable
interest since the high dipole moment (5.9~debyes) of this molecule
makes its rotational and rotovibrational transitions particularly
strong (see e.g. Zemke \& Stwalley~1980, Bellini et al.~1994, Gianturco
et al.~1996).  For this reason, the possibility of detecting LiH
features in the spectrum of the CBR and in high redshift protogalactic
clouds has raised a considerable interest both theoretically
(Dubrovich~1994, Maoli~1994, Maoli et al.~1994, Maoli et al.~1996) and
observationally (de Bernardis et al.~1993).  Specific observational
programs are in progress (Signore et al.~1994).

It is important to remind that the abundance of LiH in the pre-galactic
gas is proportional to the primordial abundance of Li (hereafter
[Li/H]$_{\rm p}$), which is still highly controversial.  It was
originally shown by Spite \& Spite~(1982) that the Li abundance on the
surface of very metal-poor Pop II stars ([Li/H]$_{\rm Pop~II}\simeq
10^{-10}$) is basically independent on metallicity (the so-called Spite
plateau).  The Pop~II value compares well with the abundances of the
other light elements predicted by standard big-bang nucleosynthesis and
might therefore represent the true primordial Li:  [Li/H]$_{\rm
p}\simeq$ [Li/H]$_{\rm Pop~II}\simeq 2\times 10^{-10}$.  However,
evidence is accumulating for a higher primordial value, of the order of
[Li/H]$_{\rm p}\simeq 10^{-9}$. From observations made with the Keck
telescope of a sample of sub-giants in the globular cluster M92,
Deliyannis, Boesgaard \& King~(1995) have recently found evidence of
significant dispersion in the measured values of [Li/H]$_{\rm Pop~II}$
values, strongly suggesting that some amount of depletion has occurred
among Spite's plateau stars. If this were the case, then the Li
abundance of Pop~II stars could be the result of depletion from a
higher initial primordial value, a result consistent with the rotating
stellar models by Pinsonneault, Deliyannis \& Demarque~(1992).  As a
possible way to distinguish between the high-[Li]$_{\rm p}$ and the
low-[Li]$_{\rm p}$ scenarios, Steigman~(1996) has recently proposed to
determine the ratio of [Li/K] from absorption features along different
lines-of-sight in the Galaxy and the Large Magellanic Cloud. At the
present stage, available data along the line-of-sight of SN1987A favor
a primordial abundance of Li lower by at least a factor $\sim 2$ than
the present Pop~I value ([Li/H]$_{\rm Pop~I}\simeq 1.6\times
10^{-9}$).

Since the main quantities determined in this study (abundances, optical
depths and anisotropies) depend on the assumed initial Li abundance, we
should expect an overall increase of their values by an order of
magnitude, if the ``high'' value of [Li]$_{\rm p}$ were to be confirmed
by future observations.

\section{Formation and destruction of Lithium Hydride}

\begin{table*}
\caption[]{List of considered reactions}
\begin{flushleft}
\begin{tabular}{lllll}
\hline
   &  reaction   &   rate (cm$^3$~s$^{-1}$)   & notes &  reference \\
\hline
1) & Li$^+$ + e $\rightarrow$ Li + $h\nu$           & 
       $1.036\times 10^{-11}[\sqrt{T_{\rm g}/107.7}\times$ & & \\
   &  &  $(1+\sqrt{T_{\rm g}/107.7})^{0.6612}\times$ & & \\
   &  &  $(1+\sqrt{T_{\rm g}/1.177\times 10^7})^{1.3388}]^{-1}$ 
   & quantal calculation & (a) \\
2) & Li + $h\nu$ $\rightarrow$ Li$^+$ + e           & 
      & detailed balance applied to (1) &  \\
3) & Li$(^2S)$ + H $\rightarrow$ LiH$(X^1\Sigma^+)$ + $h\nu$ & 
        $10^{-17}$  & semiclassical estimate & (b) \\
   &  & $1.0\times 10^{-16}T_{\rm g}^{-0.461}$ & quantal calculation & (c) \\
   &  & $(5.6\times 10^{19}T_{\rm g}^{-0.15}+7.2\times 10^{15}T_{\rm g}^{1.21})^{-1}$
   & quantal calculation (adopted) & (d),(e) \\
4) & LiH$(X^1\Sigma^+)$ + $h\nu$ $\rightarrow$ Li$(^2S)$ + H   &  
      & detailed balance applied to (3) &  \\
5) & Li$(^2P)$ + H $\rightarrow$ LiH$(X^1\Sigma^+)$ + $h\nu$ & 
      $2.6\times 10^{-16}T_{\rm g}^{0.12}\exp(-T_{\rm g}/6500)$ & 
      $A^1\Sigma^+\rightarrow X^1\Sigma^+$, quantal calc. & (f) \\
6) & LiH$(X^1\Sigma^+)$ $\rightarrow$ Li$(^2P)$ + H + $h\nu$ & 
      & detailed balance applied to (5) &  \\
7) & Li$(^2P)$ + H $\rightarrow$ LiH$(X^1\Sigma^+)$ + $h\nu$ & 
      $1.9\times 10^{-14}T_{\rm g}^{-0.34}$ & 
      $B^1\Pi\rightarrow X^1\Sigma^+$, quantal calc. & (f) \\
8) & LiH$(X^1\Sigma^+)$ $\rightarrow$ Li$(^2P)$ + H + $h\nu$ & 
      & detailed balance applied to (7) &  \\
9) & Li$^+$ + H$^-$ $\rightarrow$ Li + H &
      $6.3\times 10^{-6}T_{\rm g}^{-1/2}-7.6\times 10^{-9}+$ & & \\
   &  & $2.6\times 10^{-10}T_{\rm g}^{1/2}$ & experimental cross section & (g) \\
10) & Li + e $\rightarrow$ Li$^-$ + $h\nu$   &
      $5.7\times 10^{-17}T_{\rm g}^{0.59}\exp(-T_{\rm g}/17200)$ 
      & experimental cross section & (h) \\
11) & Li$^-$ + $h\nu$ $\rightarrow$ Li + e  &
      & detailed balance applied to (10) &  \\
12) & Li + H$^+$ $\rightarrow$ Li$^+$ + H               & 
      $2.5\times 10^{-40}T_{\rm g}^{7.9}\exp(-T_{\rm g}/1210)$ 
    & quantal calculation & (i) \\
13) & Li + H$^+$ $\rightarrow$ Li$^+$ + H + $h\nu$      &
      $1.7\times 10^{-13}T_{\rm g}^{-0.051}\exp(-T_{\rm g}/282000)$ 
      & quantal calculation & (j) \\
14) & Li$^-$ + H$^+$ $\rightarrow$ Li + H               &
      & same as (9) & (g) \\
15) & Li$^+$ + H $\rightarrow$ LiH$^+$ + $h\nu$         &
      $1.4\times 10^{-20}T_{\rm g}^{-0.9}\exp(-T_{\rm g}/7000)$ 
    & quantal calculation & (d),(e) \\
16) & LiH$^+$ + $h\nu$ $\rightarrow$ Li$^+$ + H & 
    & detailed balance applied to (15) &  \\
17) & LiH$^+$ + e $\rightarrow$ Li + H                 & 
      $3.8\times 10^{-7}T_{\rm g}^{-0.47}$ & estimate  & (k) \\
18) & LiH$^+$ + H $\rightarrow$ Li + H$_2^+$           &
      $9.0\times 10^{-10}\exp(-66400/T_{\rm g})$ & estimate & (k) \\
19) & LiH$^+$ + H $\rightarrow$ Li$^+$ + H$_2$         &
      $3.0\times 10^{-10}$ & estimate & (k) \\
20) & LiH$^+$ + H $\rightarrow$ LiH + H$^+$            &
      $1.0\times 10^{-11}\exp(-67900/T_{\rm g})$ & estimate & (k) \\
21) & Li + H$^-$ $\rightarrow$ LiH + e                 & 
      $4.0\times 10^{-10}$ & estimate & (k) \\
22) & Li$^-$ + H $\rightarrow$ LiH + e                 &
      $4.0\times 10^{-10}$ & estimate & (k) \\
23) & LiH + H$^+$ $\rightarrow$ LiH$^+$ + H            &
      $1.0\times 10^{-9}$ & estimate & (k) \\
24) & LiH + H $\rightarrow$ Li + H$_2$                 &
      $2.0\times 10^{-11}$ & estimate & (k) \\
25) & LiH + H$^+$ $\rightarrow$ LiH$^+$ + H            &
      $1.0\times 10^{-9}$ & estimate & (k) \\
26) & LiH + H$^+$ $\rightarrow$ Li$^+$ + H$_2$         &
      $1.0\times 10^{-9}$ & estimate & (k) \\
27) & LiH + H$^+$ $\rightarrow$ Li$^+$ + H$_2^+$       &
      $1.0\times 10^{-9}$ & estimate & (k) \\
\hline
\end{tabular}

\vspace{1em}

\small{(a)~Verner \& Ferland~(1996)}; (b)~Lepp \& Shull~(1984); 
(c)~Khersonskii \& Lipovka~(1993); (d)~Dalgarno, Kirby \& Stancil~(1996); 
(e)~Gianturco \& Gori Giorgi~(1996a);
(f)~Gianturco \& Gori Giorgi~(1996b); (g)~Peart \& Hayton~(1994);
(h)~Ramsbottom, Bell \& Berrington~(1994); (i)~Kimura, Dutta \& Shimakura~(1994);
(j)~Stancil \& Zygelman~(1996)
(k)~Stancil, Lepp \& Dalgarno~(1996).
\end{flushleft}
\end{table*}
 
The LiH molecule has been the subject of a large number of theoretical
investigations aimed at determining the conditions for its formation in
various astrophysical contexts.  Kirby \& Dalgarno~(1978) (see also
Bochkarev \& Khersonkii~1985) considered the photodissociation of LiH
following absorption from the $v=0$ vibrational level of the ground
$X^1\Sigma^+$ state into the vibrational continuum of the $A^1\Sigma^+$
or $B^1\Pi$ excited states.  The cross sections and rates for the
inverse radiative association processes
$$
{\rm Li}(^2P)+{\rm H}\rightarrow {\rm LiH}(A^1\Sigma^+) 
\rightarrow {\rm LiH}(X^1\Sigma^+) + h\nu,
$$
$$
{\rm Li}(^2P)+{\rm H}\rightarrow {\rm LiH}(B^1\Pi) 
\rightarrow {\rm LiH}(X^1\Sigma^+) + h\nu,
$$
have been recently computed by Gianturco \& Gori Giorgi~(1996b).  Since
the Li$(^2P)$ level lies about 1.85~eV above the Li$(^2S)$ level, the
formation of LiH molecules via electronically excited levels is
important only when the temperature is greater than some thousand
degrees K.  At lower temperatures, processes involving transitions
between vibrational levels of the ground $X^1\Sigma^+$ state and its
continuum become the dominant mechanisms of radiative association and
dissociation of LiH.  Such physical conditions are realized in the
interior of dense clouds, and in the primordial pre-galactic gas at
epoch $z\la 500$.

For the direct radiative association of Li$(^2S)$ and H,
$$
{\rm Li}(^2S)+{\rm H}\rightarrow {\rm LiH}(X^1\Sigma^+) + h\nu,
$$
Lepp \& Shull~(1984), using semi-classical arguments, estimated a rate
coefficient $R_{\rm ass}\simeq 10^{-17}$~cm$^3$~s$^{-1}$, a value
adopted in several studies of the chemistry of the early Universe (e.g.
Lepp \& Shull~1984; Puy et al.~1993; Palla et al.~1995).  Khersonskii
\& Lipovka~(1993) performed a fully quantum mechanical calculation
analysis of this reaction.  Their results show a rapid increase of the
cross sections for radiative association with the vibrational quantum
number $v$: the molecule is formed preferencially in vibrational levels
with $v=$15--18.  The resulting rate coefficient for the direct
radiative association of LiH is in good agreement with the
semi-classical estimate of Lepp \& Shull~(1984) (see Table~1).

More recently, the LiH chemistry in the primordial gas has been
reanalyzed in detail by Dalgarno, Kirby \& Stancil~(1996) and by
Stancil, Lepp \& Dalgarno~(1996, hereafter SLD). The accurate quantal
calculation of the radiative association rate of LiH performed by
Dalgarno et al.~(1996) gives a value of $R_{\rm ass}$ smaller than the
one computed by Lepp \& Shull~(1984) and Khersonskii \& Lipovka~(1993)
by two--three orders of magnitude. The reason for such a large
difference is not clear, but it may be related to the approximations
adopted by the latter authors. The quantal calcualation of $R_{\rm
ass}$ for reaction (4) recently performed by Gianturco \& Gori
Giorgi~(1996a) has confirmed SLD's results to a very high degree of
precision. This is even more remarkable, since the two calculations are
based on independent sets of potential curves and wavefunctions.

The total rate of radiative association of LiH is represented
graphically in Fig.~1 as function of the temperature $T_{\rm g}$ for a
gas where the relative populations of the $^2S$ and $^2P$ levels of Li
are assumed to be in thermal equilibrium at temperature $T_{\rm g}$.
It is clear from the Figure that radiative association via excited
electronic levels is the dominant formation process of LiH when
the gas temperature is larger than $\sim 2000$~K. Otherwise, LiH is
formed only by direct radiative association.

\section{Chemical evolution of the primordial gas}


\begin{table*}
\caption[]{Final fractional abundances}
\begin{flushleft}
\begin{tabular}{lllllllll}
\hline
  & \multicolumn{2}{l}{$\eta_{10}=0.7$} &  & \multicolumn{2}{l}{$\eta_{10}=4.5$} 
				        &  & \multicolumn{2}{l}{$\eta_{10}=10$} \\ 
\cline{2-3} \cline{5-6} \cline{8-9}
         &  $z=10$   &  $z=0$  & &  $z=10$   &  $z=0$  & &  $z=10$   &  $z=0$  \\
\hline
H$^+$, e & $4.6\times 10^{-3}$ & $4.6\times 10^{-3}$ & & $7.2\times 10^{-4}$ 
	 & $7.1\times 10^{-4}$ & & $3.2\times 10^{-4}$ & $3.2\times 10^{-4}$ \\
H$^-$ & $1.9\times 10^{-11}$ & $1.9\times 10^{-13}$ & & $1.5\times 10^{-12}$ 
       & $1.5\times 10^{-14}$ & & $4.9\times 10^{-13}$ & $4.9\times 10^{-15}$ \\
Li     & $8.2\times 10^{-10}$ & $8.5\times 10^{-10}$ & & $1.6\times 10^{-10}$ 
       & $1.7\times 10^{-10}$ & & $8.9\times 10^{-10}$ & $9.3\times 10^{-10}$ \\
Li$^+$ & $2.8\times 10^{-10}$ & $2.5\times 10^{-10}$ & & $4.5\times 10^{-11}$ 
       & $3.8\times 10^{-11}$ & & $2.1\times 10^{-10}$ & $1.7\times 10^{-10}$ \\
Li$^-$ & $5.0\times 10^{-20}$ & $4.2\times 10^{-22}$ & & $4.3\times 10^{-21}$ 
       & $3.9\times 10^{-23}$ & & $1.6\times 10^{-20}$ & $1.5\times 10^{-22}$ \\
LiH    & $1.5\times 10^{-18}$ & $4.2\times 10^{-19}$ & & $2.2\times 10^{-19}$ 
       & $9.8\times 10^{-20}$ & & $1.2\times 10^{-18}$ & $4.8\times 10^{-19}$ \\
LiH$^+$ & $8.1\times 10^{-17}$ & $1.2\times 10^{-16}$ & & $8.5\times 10^{-18}$ 
	& $2.2\times 10^{-17}$ & & $2.9\times 10^{-17}$ & $1.0\times 10^{-16}$ \\
\hline
\end{tabular}
\end{flushleft}
\end{table*}

\begin{figure} 
\centerline{\psfig{figure=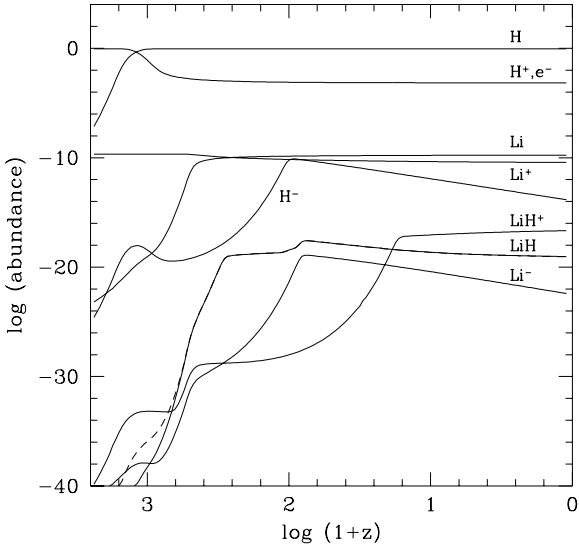,width=8.4cm}}
\caption[]{Fractional abundance (by number) of atoms, ions and
molecules as function of the redshift for the standard model.
The effect of the radiative association of LiH in excited electronic
levels is shown by the {\em dashed} line.}
\end{figure}

\begin{figure}
\centerline{\psfig{figure=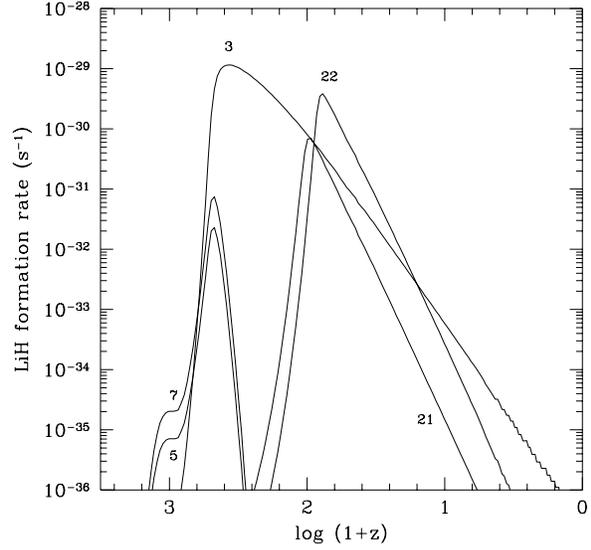,width=8.4cm}}
\caption[]{Processing rates of the most important reactions for the 
formation of LiH. The curves are labelled with the corresponding 
reaction numbers listed in Table~1
} 
\end{figure}

We consider now the chemical evolution of the pregalactic gas in the
framework of Friedmann cosmological models.  Our standard model is
characterized by a Hubble constant $H_0=67$~km~s$^{-1}$~Mpc$^{-1}$ (van
den Bergh~1989), a closure parameter $\Omega_0=1$, a baryon-to-photon
ratio $\eta_{10}=4.5$ (Galli et al.~1995), and a present radiation
temperature $T_0=2.726$~K (Mather et al.~1993). The initial fractional
abundance of Li nuclei for this model is $2.1\times 10^{-10}$ (Smith,
Kawano \& Malaney~1993). The integration of the chemical network starts
at $z=2500$, with H and Li fully ionized and He fully recombined.  We
have also considered the possibility of a high [Li/H]$_{\rm p}$, as
discussed in the Introduction. If [Li/H]$_{\rm p}$ is close to
$10^{-9}$, standard big-bang nucleosynthesis allows either a ``low'' or
a ``high'' value of $\eta_{10}$ ($\eta_{10}\simeq 0.7$ or
$\eta_{10}\simeq 10$, respectively). The primordial Li abundance
predicted by models of inhomogeneous Big Bang nucleosynthesis can even
be considerably larger than the values adopted here, but since its
value depends sensitively on a number of parameters and assumptions
(see e.g. Mathews et al.~1990), for simplicity we consider models based
only on standard Big Bang nucleosynthesis results. The initial
fractional abundances of H, He and Li are 0.936, 0.064 and $1.1\times
10^{-9}$ for the $\eta=0.7$ model, 0.926, 0.074 and $2.1\times
10^{-10}$ for the $\eta=4.5$ model, and 0.923, 0.077 and $1.1\times
10^{-9}$ for the $\eta=10$ model.

Our chemical network includes about one hundred reactions among 21
chemical species.  Reactions involving Li-bearing species and the
adopted rates are listed in Table~1. For the recombination rate
(reaction 1) we have adopted the accurate fitting formula given by
Verner \& Ferland~(1996), based on the photoionization cross sections
computed by Li by Peach, Saraph \& Seaton~(1988) and Gould~(1978). The
recombination rate coefficient of Li shown in Table~1 is in good
agreement with the results by Caves \& Dalgarno~(1978).

The evolution of the gas temperature $T_{\rm g}$ is governed by the equation
(see e.g. Galli~1990; Puy et al.~1993; Palla et al.~1995)
\beq
\label{tempe}
\frac{{\rm d}T_{\rm g}}{{\rm d}t}=-2T_{\rm g}\frac{\dot{R}}{R}+
\frac{2}{3kn}[(\Gamma-\Lambda)_{\rm Compton}+(\Gamma-\Lambda)_{\rm mol}].
\enq
The first term represents the adiabatic cooling associated to the
expansion of the Universe ($R$ is the scale factor of the Universe).
The other two terms represent respectively the net transfer of energy
from the CBR to gas (per unit time and unit volume) via Compton
scattering of CBR photons on electrons,
\beq
(\Gamma-\Lambda)_{\rm Compton}=\frac{4k\sigma_{\rm T}aT_{\rm r}^4
(T_{\rm r}-T_{\rm g})}{m_{\rm e}c}n_{\rm e},
\enq
and via excitation and de-excitation of molecular transitions,
\beq
(\Gamma-\Lambda)_{\rm mol}=\sum_{i>j}(n_iC_{ij}-n_jC_{ji})h\nu_{ij},
\enq
where $C_{ij}$ and $C_{ji}$ are the collisional excitation and
de-excitation coefficients and $n_{i}$ are the level populations.  The
adopted expressions for the radiative and collisional coefficients of
LiH as well as the procedure followed to determine the level
populations are summarized in Appendix B.  For the molecular heating
and cooling of the gas we have considered the contributions of H$_2$, HD
and LiH.

When the rate of collisional de-excitation of the roto-vibrational
molecular levels is faster (slower) than their radiative decay, the
energy transfer function $(\Gamma-\Lambda)_{\rm mol}$ can become an
effective heating (cooling) source for the gas (cf.  Khersonskii~1986,
Puy et al.~1993). In general, the contribution of molecules to the
heating/cooling of the gas becomes important only during the phase of
collapse of protogalactic objects (see Sect.~4).

The chemical/thermal network is completed by the equation for the redshift
\beq
\frac{{\rm d}t}{{\rm d}z}=-\frac{1}{H_0(1+z)^2\sqrt{1+\Omega_0z}},
\enq
where $H_0$ is the Hubble constant.  The density $n(z)$ of baryons 
at redshift $z$ is 
\beq
n(z)=\Omega_{\rm b}n_{\rm cr}(1+z)^3,
\enq
where $n_{\rm cr}$ is the critical density. The abundance of atoms and
molecules is measured with respect to the total number of baryons:
$f({\rm H})=n({\rm H})/n$ and so on.

\subsection{Results}

The evolution of the fractional abundances of H, H$^+$, H$^-$, Li,
Li$^+$, Li$^-$, LiH$^+$ and LiH for the standard model is shown in
Fig.~2 as function of the redshift $z$. The abundances at $z=10$ and
$z=0$ for all the models considered are listed in Table~2.

In all models, the final abundance of LiH is extremely small, of the
order of $10^{-19}$--$10^{-18}$. The maximum final abundance ($\sim
5\times 10^{-19}$) is obtained for the $\eta_{10}=10$ model.  Fig.~3
shows the relative contribution of individual chemical reactions to the
formation of LiH in the early Universe.  Radiative association from
Li($^2P$) is significant only for $z>650$, with the process
$B^1\Pi\rightarrow X^1\Sigma^+$ being dominant. Otherwise, LiH is
formed mostly by radiative association from Li($^2S$) (reaction 3). At
$z\sim 80$ the reactions of associative detachment (21) and, most
importantly, (22) also contribute to the formation of LiH. Notice
however that although the value of the rate coefficients for the latter
reactions is rather uncertain (see SLD), the asymptotic value of the
abundance of LiH is largely determined by reaction (3), whose rate is
now well established (Dalgarno, Kirby \& Stancil~1996, Gianturco \&
Gori Giorgi~1996a).

The evolution of Li$^-$ for $z< 100$ closely mimicks that of H$^-$,
both ions being removed at low temperature by reactions of mutual
neutralization with H$^+$ at a rate proportional to $T_{\rm
g}^{-1/2}$.  Since $T_{\rm g}\propto (1+z)^2$ in this range of reshift,
the abundances of H$^-$ and Li$^-$ decrease like $(1+z)$, as shown in
Fig.~2.  Our computed abundances of H$^+$, H$^-$, Li and LiH$^+$ at $z=10$
are in excellent agreement with the results obtained by SLD (within
10--20\%), our abundance of Li$^+$ being a factor 1.8 smaller. Our values
for Li$^-$ and  LiH at $z=10$ are smaller by large factors ($\sim 900$
and $\sim 300$, respectively) than those published by SLD, but the two
computations give essentially the same results when a sign mistake in
the rate adopted by SLD for reaction (10) is corrected.

An interesting result is that LiH$^+$ is more abundant than LiH for $z
\la 18$, by a factor $\sim 200$ for the standard model. This is in
agreement with the early prediction by Dalgarno \& Lepp~(1987).

\begin{figure}
\centerline{\psfig{figure=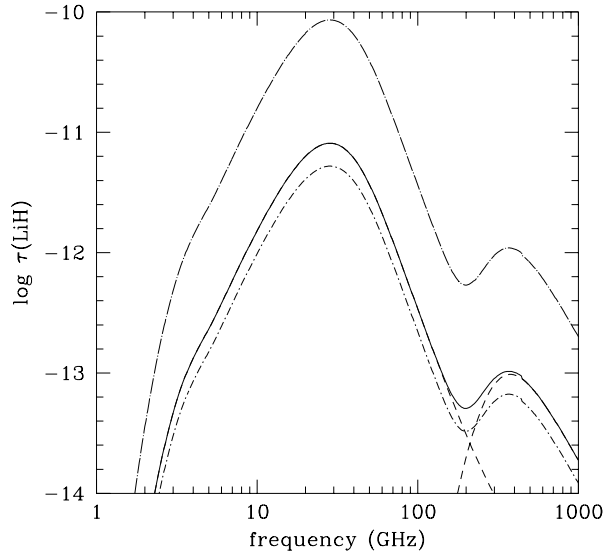,width=8.4cm}}
\caption[]{
Optical depth of the Universe generated by Thomson scattering of CBR
photons on LiH molecules for the standard model ({\em solid} curve) and
for the high primordial Li cases ({\em dot short dashed} curve,
$\eta_{10}=0.7$; {\em dot long dashed} curve, $\eta_{10}=10$). For the
standard model, the contribution of rotational and rotovibrational
transitions is shown by the {\em dashed} curves.}
\end{figure}

\subsection{Optical depth}

Having obtained the LiH abundance as function of the redshift, we are
now able to compute, with the help of the formulae listed in Appendix
A, the total optical depth of the Universe due to elastic Thomson
scattering of CBR photons on LiH molecules (see eq.~[\ref{taufin}]).
The result is shown in Fig.~4 as function of the frequency of
observation $\nu$. The maximum optical depth for the standard model is
$\tau_{\rm LiH}^{\rm max}=6\times 10^{-11}$, at $\nu^{\rm max}=
30$~GHz. Pure rotational transitions of the $v=0$ level dominate at
this frequency.  The general behaviour of $\tau_{\rm LiH}(\nu)$ shown
in Fig.~4 is in agreement with the one obtained by Maoli et
al.~(1994).  The attenuation of CBR anisotropies being of order
$\tau_{\rm LiH}$, little room is left for the ``molecular blurring''
suggested by Dubrovich~(1993).  Even for the high [Li/H]$_{\rm p}$
model with $\eta_{10}=10$, where a larger optical depth is obtained,
the induced attenuation is still insignificant.

\section{A simple model of a collapsing primordial cloud}

\begin{table*}
\caption[]{Physical quantities for a $10^{10}$~$M_\odot$ collapsing protogalaxy}
\begin{flushleft}
\begin{tabular}{lrlllllllll}
\hline
$t/t_{\rm ff}$ & $z$ & $n$ & $T_{\rm g}$ & $R$ & $\theta$ & $v_{\rm coll}$ 
	       & $v_{\rm pec}$ & $f({\rm H}_2)$ & $f({\rm HD})$ 
	       & $f({\rm LiH})$ \\
 & & (cm$^{-3}$) & (K) & (kpc) & ($\arcsec$) & (km~s$^{-1}$) & (km~s$^{-1}$) & &  \\
\hline
0.0  & 10.00 & $1.38\times 10^{-3}$ & 9.6 & 38.8 & 28.2 & 0.0 & 90  
& $4.5\times 10^{-6}$ & $5.3\times 10^{-9}$ & $2.2\times 10^{-19}$ \\
0.1  & 7.45  & $1.41\times 10^{-3}$ & 9.8 & 38.5 & 22.8 & 3.8 & 103 
& $4.5\times 10^{-6}$ & $5.4\times 10^{-9}$ & $2.4\times 10^{-19}$ \\
0.2  & 6.03  & $1.49\times 10^{-3}$ & 10  & 37.8 & 19.6 & 7.6 & 113 
& $4.6\times 10^{-6}$ & $5.4\times 10^{-9}$ & $2.8\times 10^{-19}$ \\
0.3  & 5.10  & $1.64\times 10^{-3}$ & 11  & 36.6 & 17.2 & 12  & 121 
& $4.6\times 10^{-6}$ & $5.4\times 10^{-9}$ & $3.0\times 10^{-19}$ \\
0.4  & 4.41  & $1.91\times 10^{-3}$ & 12  & 34.8 & 15.2 & 16  & 129 
& $4.6\times 10^{-6}$ & $5.4\times 10^{-9}$ & $3.4\times 10^{-19}$ \\
0.5  & 3.91  & $2.34\times 10^{-3}$ & 14  & 32.5 & 13.4 & 21  & 135 
& $4.7\times 10^{-6}$ & $5.5\times 10^{-9}$ & $3.6\times 10^{-19}$ \\
0.6  & 3.47  & $3.22\times 10^{-3}$ & 17  & 29.2 & 11.4 & 27  & 142 
& $4.7\times 10^{-6}$ & $5.6\times 10^{-9}$ & $4.2\times 10^{-19}$ \\
0.7  & 3.13  & $4.99\times 10^{-3}$ & 23  & 25.3 & 9.4  & 33  & 148 
& $4.8\times 10^{-6}$ & $5.8\times 10^{-9}$ & $4.8\times 10^{-19}$ \\
0.8  & 2.86  & $9.81\times 10^{-3}$ & 36  & 20.2 & 7.2  & 45  & 153 
& $5.0\times 10^{-6}$ & $6.5\times 10^{-9}$ & $5.6\times 10^{-19}$ \\
0.9  & 2.63  & $3.31\times 10^{-2}$ & 80  & 13.4 & 4.8  & 65  & 157 
& $5.8\times 10^{-6}$ & $9.5\times 10^{-9}$ & $7.2\times 10^{-19}$ \\
0.95 & 2.54  & $1.05\times 10^{-1}$ & 172 & 9.15 & 3.2  & 85  & 159 
& $7.9\times 10^{-6}$ & $1.1\times 10^{-8}$ & $9.2\times 10^{-19}$ \\
\hline
\end{tabular}
\end{flushleft}
\end{table*}

So far our attention has been directed exclusively to the global
effects of LiH molecules in the {\em homogeneous} primordial gas. One
of the main limitations to the occurence of detectable features is the
low molecular abundance obtained in such a context. It is therefore of
interest to follow the chemical (and thermal) evolution of the first
density inhomogeneities that emerged from the expanding primordial
gas.  Those initial density perturbations whose mass exceeded the Jeans
mass were unstable to gravitational collapse and evolved to form
galaxies and galaxy clusters. The increase in density during the first
(quasi adiabatic) phase of collapse can provide suitable conditions for
the build-up of considerably larger molecular fraction than in the
surrounding expanding Universe. In addition, velocity gradients in a
cloud may contribute to enhance the effective optical depth of
localized regions in the Universe (Maoli et al.~1994, 1996).

Let us consider then the collapse of a purely baryonic primordial 
cloud in order to determine the abundance of the relevant molecular
species as function of time and to predict the level of anisotropies
induced by molecular transitions during the collapse of the cloud.  We
follow the approach of Lahav~(1986) considering a spherical homogeneous
cloud of mass $M$, uniform density $n$ and radius $R$ contracting (or
expanding) in a homologous fashion ($v(r)\propto r$). We assume that
the radius of the cloud reaches its maximum value at the {\em
turn-around} epoch $\zta$, when the density and temperature of the
cloud are related to the corresponding quantities in the ambient gas by
\beq
\rho_{\rm ta}=\left(\frac{3\pi}{4}\right)^2\rho_{\rm g}(\zta),
\enq
\beq
T_{\rm ta}=\left(\frac{3\pi}{4}\right)^{4/3}T_{\rm g}(\zta).
\enq
The radius at turn around $R_{\rm ta}$ is fixed by the mass and density
of the cloud. In adiabatic collapse the temperature of
the gas in the cloud, $T_{\rm g}$, increases like $R^{-2}$.

By virtue of the hypothesis of homologous collapse, the primordial
cloud may be treated in the same way as the expanding Universe in the
previous Sections. For example, eq.~(\ref{tempe}) for the gas
temperature still holds if one replaces the scale factor of the
Universe with the cloud's radius.  The optical depth of the cloud by
elastic Thomson scattering of CBR photons on LiH molecules is given
again by eq.~(\ref{taufin}), but the cosmological velocity gradient is
now replaced by the velocity gradient due to gravitational collapse. A
more sophisticated calculation of the optical depth, taking into
account projection effects, can be found in Maoli et al.~(1996).

Defining a non-dimensional radius $\xi$ and temperature $\theta$ by
\beq
R=\xi R_{\rm ta},~~~~T_{\rm g}=T_{\rm ta}\theta \xi^{-2}.
\enq
the equations for the radius and for the temperature can be written as
(Lahav~1986)
\beq
\label{xi}
\frac{{\rm d}^2\xi}{{\rm d}t^2}=\frac{5kT_{\rm ta}}{\mu m_H R_{\rm ta}^2}
\theta\xi^{-3}-\frac{GM}{R_{\rm ta}^3}\xi^{-2},
\enq
and
\beq
\label{theta}
\frac{{\rm d}\theta}{{\rm d}t}= \frac{2}{3}\frac{(\Gamma-\Lambda)_{\rm Compton}+
(\Gamma-\Lambda)_{\rm mol}}{nkT_{\rm ta}}\xi^2,
\enq
It is convenient to express time in units of the free-fall time
for the cloud,
\beq
t_{\rm ff}=\pi\left(\frac{R_{\rm ta}^3}{8GM}\right)^{1/2}.
\enq
Once the mass $M$ and the turn-around redshift $z_{\rm ta}$ are fixed,
the evolution of the model cloud is determined by the
integration of eq.~(\ref{xi}) and (\ref{theta}). A relation between $M$
and $z_{\rm ta}$ can be obtained by prescribing the power spectrum of
the density perturbations (see e.g. Gunn \& Gott~1972, Lahav~1986). For
simplicity, we prefer to use a reference value $z_{\rm ta}=10$ for all
masses.

 
\begin{figure}
\centerline{\psfig{figure=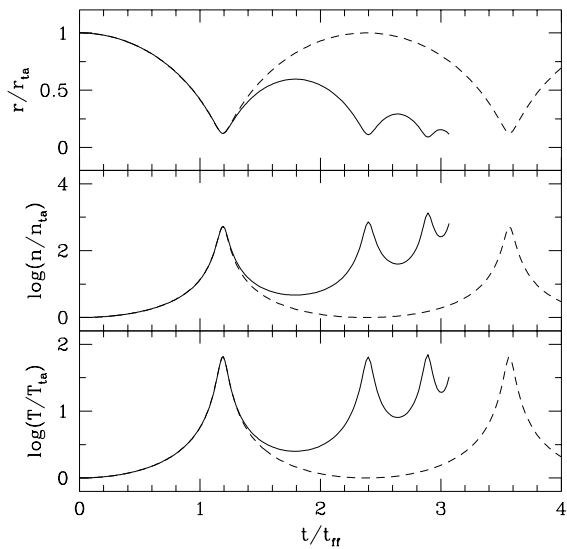,width=8.4cm}}
\caption[]{Time evolution of radius, density and temperature for a primordial cloud 
of mass $M=5\times 10^5$~$M_\odot$. All quantities are normalized to their initial
turn around values. Time is in units of the free-fall time. {\em Dashed lines:}
adiabatic evolution; {\em solid lines:} non-adiabatic evolution.}
\end{figure}

\begin{figure}
\centerline{\psfig{figure=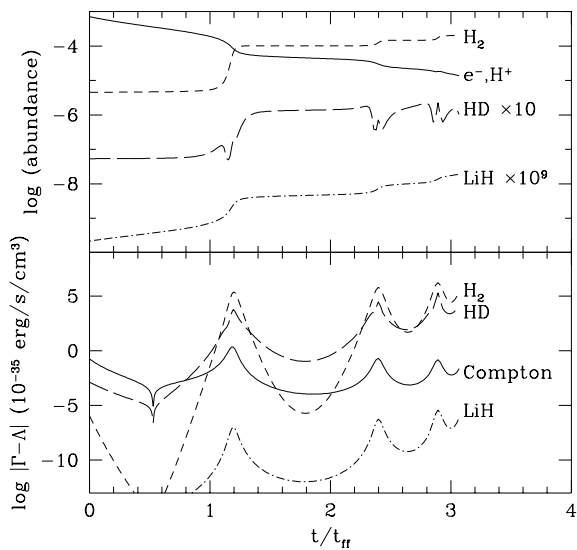,width=8.4cm}}
\caption[]{{\em Upper panel:} fractional abundances of the relevant atomic and 
molecular species as function of time for the same primordial cloud as in Fig.~5.
{\em Lower panel:} absolute value of the net energy transfer due to molecular and 
Compton heating/cooling processes.}
\end{figure}

We consider here a low-mass cloud with
$M=5 \times 10^5$~$M_\odot$, slightly above the Jeans mass at $z_{\rm
ta}=10$ ($M_{\rm J}= 3\times 10^5$~$M_\odot$), and a galactic-size
cloud with $M=10^{10}$~$M_\odot$.  Fig.~5 shows the time evolution of
the cloud's radius, density and temperature (all normalized to ther
initial turn-around values) for the $M=5\times 10^5$~$M_\odot$ case.
The evolution is followed for about three free-fall times for
illustrative purposes only; in a more realistic model, relaxing the
assumption of homologous contraction/expansion, the actual evolution of
the cloud is expected to deviate from the one shown here owing to the
shocking of different mass shells (see e.g. Haiman, Thoul \&
Loeb~1996). Because of molecular cooling, the model cloud performs a
series of oscillations of decreasing amplitude and period, in striking
contrast with the corresponding evolution in the adiabatic case (dashed
lines).  Fig.~6 shows the fractional abundance of the relevant atomic
and molecular species and the net energy exchange between gas and
radiation due to molecular and Compton processes, for the same case as
in Fig.~5.  Both processes act as a net heating for the cloud for
$t<0.53\;t_{\rm ff}$, when $T_{\rm r} > T_{\rm g}$, and as a net
cooling afterwards. It is important to notice the relevant role played
by HD molecules, in addition to H$_2$, for cooling the gas during the
early phases of collapse. As shown in Fig.~6, heating/cooling by LiH
molecules is negligible at all times.

The relevant physical characteristics during the collapse of the
$M=10^{10}$~$M_\odot$ cloud are summarized in Table 3. The listed
quantities are:  redshift $z$, cloud density $n$, temperature $T_{\rm
g}$, radius $R$, angular diameter $\theta$, collapse velocity $v_{\rm
coll}$ of the outermost shell, peculiar velocity $v_{\rm pec}$, and
fractional abundances of the most relevant molecular species.
The angular diameter $\theta$ of a spherical cloud of proper 
radius $R$ at redshift $z$ is computed from the formula (see e.g. Peebles~1971)
\beq
\theta(R,z)=\frac{H_0 R}{c}
\frac{\Omega_0^2(1+z)^2}{\Omega_0z+(\Omega_0-2)(\sqrt{1+\Omega_0z}-1)}.
\enq
For the peculiar velocity we follow Maoli et al.~(1994) assuming the simple law 
of evolution of velocity dispersion in the linear regime (see e.g. Peebles~1980)
\beq
v_{\rm pec}(z)=\frac{v_{\rm pec}(0)}{\sqrt{1+z}},
\enq
where $v_{\rm pec}(0)=300$~km~s$^{-1}$ is the present-day galaxy
velocity dispersion (Peebles~1993). We assume that this velocity is
directed along the line of sight.  Table 4 shows the value of the
optical depth of the three lowest rotational transitions of LiH as
function of time, computed according to eq.~(\ref{taufin}).

Anisotropies of the CBR generated by the peculiar motion of the
protogalactic cloud can be positive or negative depending on whether
the cloud is approaching of receding (see eq.~[\ref{sz}]). Although the
total peculiar velocity of a collapsing cloud results from the
combination of $v_{\rm pec}$ and the collapse velocity $v_{\rm coll}$,
we see from the values listed in Table~3 that the latter can be
neglected for most of the collapse phase.  For a perturbation evolving
in a homologous way, the instant of turn around, when the radial
collapse velocity exactly compensates the overall cosmological
expansion, presents interesting peculiarities. At this time, in fact,
all the molecules in the cloud are ``at rest'', and they all contribute
to the optical depth.  Similar considerations hold for the line width:
at the redshift of turn-around the width of a line is the thinnest,
practically equal to the thermal width $(\Delta\nu/\nu)_{\rm th}$.
This effect was originally predicted by Zeldovich~(1978), who pointed
out that weak and narrow spectral features generated in protogalaxies
(or protoclusters) reaching their radii of turn-around could be
superimposed to the spectrum of the CBR.

\section{Discussion and Conclusions}

The presence of LiH in collapsing protogalactic clouds may induce
fluctuations of the CBR brightness temperature on scales of $\sim
10\arcsec$ which can be in principle revealed by the use of arrays of
telescopes and aperture synthesis techniques.  However, it is clear
from the values of $\tau_{\rm LiH}$ listed in Table~4 that the expected
spectral features of LiH in protoclouds are extremely weak:  the
associated anisotropies in the CBR temperature, according to
eq.~(\ref{sz}) are completely negligible during most of the collapse
phase. Although the optical depth rises very rapidly towards the end of
the collapse phase, the resulting value of $\Delta T/T$ cannot be
predicted quantitatively on the basis of the simple model adopted in
this work, and a more realistic approach should be taken.

The relation between the angular size $\theta$ of the model cloud and
the redshifted frequencies of the three lowest rotational transitions
of LiH is shown in graphic form in Fig.~7, where the coverage of the
plane $\nu$--$\theta$ of current telescopes is also schematically
displayed. Probably the most appealing observational prospective is to
search for CBR temperature fluctuations associated with the $J=1$--0
rotational transition of LiH in ``evolved'' protogalaxies with the IRAM
Plateau-de-Bure Interferometer at 3~mm.  Although the actual level of
temperature fluctuations is hard to predict quantitatively for such
evolved objects, it might be close to the sensitivity limit of this
instrument.  It is also important to remind that the values of $\Delta
T/T$ obtained above are proportional to the assumed value of the
primordial abundance of Li (the results presented above are obtained
with [Li/H]$_{\rm p}=2.1\times 10^{-10}$).

\begin{figure}
\centerline{\psfig{figure=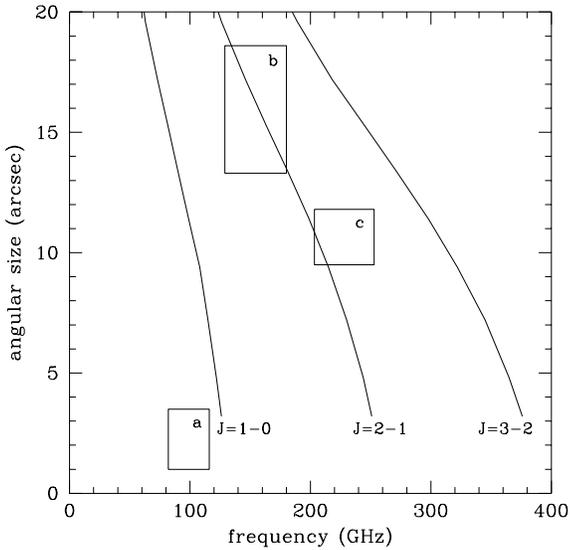,width=8.4cm}}
\caption[]{Relation between the angular size $\theta$ of a $10^{10}$~$M_\odot$ 
collapsing cloud and the redshifted frequency $\nu$ of the three lowest
rotational transitions of LiH.  Boxes represent indicatively regions of
the plane $\nu$--$\theta$ that can be explored with current teloscopes:
IRAM Plateau-de-Bure Interferometer (a), IRAM 30~m (b and c).}
\end{figure}

\begin{table}
\caption[]{Optical depth of the first three rotational transitions of
LiH for a $10^{10}$~$M_\odot$ collapsing primordial cloud}
\begin{flushleft}
\begin{tabular}{llll}
\hline
$t/t_{\rm ff}$ & $\tau_{10}$ & $\tau_{21}$ &$\tau_{32}$ \\
\hline
0.0  & $2.8\times 10^{-10}$ & $4.2\times 10^{-10}$ & $1.8\times 10^{-10}$ \\
0.1  & $1.5\times 10^{-11}$ & $1.6\times 10^{-11}$ & $4.4\times 10^{-12}$ \\
0.2  & $1.1\times 10^{-11}$ & $9.6\times 10^{-12}$ & $1.7\times 10^{-12}$ \\
0.3  & $1.0\times 10^{-11}$ & $7.0\times 10^{-12}$ & $8.6\times 10^{-13}$ \\
0.4  & $1.1\times 10^{-11}$ & $6.2\times 10^{-12}$ & $5.4\times 10^{-13}$ \\
0.5  & $1.2\times 10^{-11}$ & $5.6\times 10^{-12}$ & $3.6\times 10^{-13}$ \\
0.6  & $1.4\times 10^{-11}$ & $5.6\times 10^{-12}$ & $2.6\times 10^{-13}$ \\
0.7  & $1.9\times 10^{-11}$ & $6.4\times 10^{-12}$ & $2.2\times 10^{-13}$ \\
0.8  & $2.8\times 10^{-11}$ & $8.2\times 10^{-12}$ & $2.2\times 10^{-13}$ \\
0.9  & $5.8\times 10^{-11}$ & $1.5\times 10^{-11}$ & $3.0\times 10^{-13}$ \\
0.95 & $1.2\times 10^{-10}$ & $3.0\times 10^{-11}$ & $1.4\times 10^{-12}$ \\
\hline
\end{tabular}
\end{flushleft}
\end{table}
 
The main conclusions of our study are:

({\em a}\/) due to its low rate of radiative association, the final
abundance of LiH is very small and the molecule is only a minor
constitutent of the primordial gas. In our standard model,
characterized by $H_0=67$~km~s$^{-1}$~Mpc$^{-1}$, $\eta_{10}=4.5$,
$\Omega_0=1$, the abundance is $f({\rm LiH})\simeq 10^{-19}$ at
$z=10$.  Such a low value casts serious doubts about the use of LiH as
a probe of the physical conditions of the gas at very high redshift.
These findings reinforce the conclusions reached by SLD in a similar analysis.
The abundance of LiH$^+$ for $z\la 18$ is about two orders of magnitude
larger than the abundance of LiH.

({\em b}\/) Taking into account the possibility of a higher value of
[Li]$_{\rm p}$, the final abundance of LiH and LiH$^+$ may increase up
to $\sim 10^{-18}$--$10^{-17}$.

({\em c}\/) The estimate of the optical depth due to resonant
scattering of the CBR photons on LiH molecules yields a value of
$\tau^{\rm max}_{\rm LiH}\sim 10^{-11}$--$10^{-10}$ at a frequency of
30 GHz.  Most of the contribution to the optical depth comes from
rotational transitions beteween high $J$ levels.  The attenuation of
pre-existing anisotropies of the CBR is therefore completely
negligible.

({\em d}\/) The enhancement of the LiH abundance during the collapse of
protogalactic clouds gives rise to fluctuations in the temperature of
the CBR associated with the peculiar motion of the object.  For a
$10^{10}~M_\odot$ protogalactic cloud reaching its turn-around at
$z_{\rm ta}=10$, the expected level of anisotropy is far too low to be
of any observational relevance during most of the collapse phase. 
The actual abundance of LiH in the latest phases of collapse cannot be 
quantitatively predicted by the methods presented here.

\subsection*{Acknowledgements}
We thank Riccardo Cesaroni, Roberto Maoli and Francesco Palla for useful
comments and conversations. We wish to thank in particular Dr.~Paola Gori Giorgi 
for making her work available to us in advance of publication and for a large
number of clarifying discussions.

\appendix
\section{radiative transfer in the expanding Universe}

The equation of radiative transfer in the expanding Universe can be written as
(e.g. Peebles~1971)
\beq
\label{rt}
\frac{1}{c}\frac{{\rm d}J_\nu}{{\rm d}z}=\frac{\kappa_\nu J_\nu-j_\nu}{H_0(1+z)^2
\sqrt{1+\Omega_0z}} +\frac{3J_\nu}{c(1+z)},
\enq
where
\beq
\kappa_\nu=\frac{c^2}{8\pi\nu^2}n_lA_{ul}\frac{g_u}{g_l}
\left(1-\frac{g_ln_u}{g_un_l}\right)\phi(\nu-\nu_{ul}),
\enq
\beq
j_\nu=\frac{h\nu}{4\pi}n_uA_{ul}\phi(\nu-\nu_{ul}).
\enq
Here $\phi(\nu-\nuul)$ is the normalized line-shape function for a
transition of frequency $\nuul$.
 
In terms of the function ${\cal J}_\nu=J_\nu(1+z)^{-3}$, eq.~(\ref{rt})
takes the form
\beq
\label{stan}
\frac{H_0(1+z)^2\sqrt{1+\Omega_0z}}{c\kappa_\nu}
\frac{d{\cal J}_\nu}{dz}={\cal J}_\nu-\frac{S_\nu,}{(1+z)^3}
\enq
where $S_\nu$ is the source function defined as
\beq
S_\nu=\frac{j_\nu}{\kappa_\nu}=\frac{2h\nu^3}{c^2}
\left(\frac{g_un_l}{g_ln_u}-1\right)^{-1}.
\enq
Eq.~(\ref{stan}) can be formally solved in the usual way multiplying each term 
for the integrating factor $e^\tau$ where
\beq
\label{tau}
\tau_\nu=-\int\frac{c\kappa_\nu}{H_0(1+z)^2\sqrt{1+\Omega_0 z}}dz,
\enq
obtaining
\beq
\label{sol}
{\cal J}_\nu(\tau_\nu)={\cal J}_\nu(0)e^{-\tau_\nu}
+e^{-\tau_\nu}\int_0^{\tau_\nu}\frac{S_\nu}{(1+z)^3} 
e^{\tau_\nu^\prime}d\tau_\nu^\prime.
\enq
 
Consider now a single transition of frequency $\nuul$. Absorption of
radiation observed today at frequency $\nu$ occurs in an interval of
the order of the thermal width $\Delta\nu_{\rm th}$ around the line
frequency $\nu_{ul}$, or, equivalently, in a range of redshift
$\Delta z_{\rm i}=(1+z_{\rm i})\Delta\nu_{\rm th}/\nu_{ul}$
around the ``interaction'' redshift $z_{\rm i}$ defined by the condition
\beq
\nu(1+z_{\rm i})=\nuul.
\enq
If the Doppler width of the line is small, as it is the case in the
present context, then the size (in redshift) of the interaction region
around $z_{\rm i}$ is also small, and the integral in eq.~(\ref{tau})
can be simplified as
\beq
\label{taufin}
\tau(z_{\rm i}) = \frac{c^3}{8\pi\nuul^3}A_{ul}\frac{g_u}{g_l}
\left[1-\frac{g_ln_u(z_{\rm i})}{g_un_l(z_{\rm i})}\right]n_l(z_{\rm i}) 
\left|\frac{{\rm d}v}{{\rm d}r}\right|^{-1}_{z_{\rm i}},
\enq
where 
\beq
\left|\frac{{\rm d}v}{{\rm d}r}\right|_{z_{\rm i}}=H_0(1+z_{\rm i})\sqrt{1+\Omega_0z_{\rm i}}
\enq
is the ``cosmological'' velocity gradient, due to the expansion of the 
Universe. The region contributing to the optical depth has a size
\beq
\Delta\ell=c\frac{{\rm d}t}{{\rm d}z}\Delta z_{\rm i}=\frac{\Delta\nu_{\rm th}}{\nu_{ul}}
\frac{c}{|dv/dr|_{z_{\rm i}}}.
\enq

In the same way, since 
\beq
\int_0^\tau \frac{S_\nu}{(1+z)^3} e^{\tau^\prime} d\tau^\prime=
\frac{S_\nu(z_{\rm i})}{(1+z_{\rm i})^3}[1-e^{-\tau(z_{\rm i})}],
\enq
eq.~(\ref{sol}) gives 
\beq
J_\nu[\tau(z_{\rm i})]=J_\nu(0)e^{-\tau(z_{\rm i})}+e^{-\tau(z_{\rm i})}
[1-e^{-\tau(z_{\rm i})}]S_\nu(z_{\rm i}),
\enq
and the relative perturbation in the CBR intensity at the present epoch is
\beq
\left.\frac{\Delta J_\nu}{J_\nu}\right|_{z=0}=
[R(z_{\rm i})-1][1-e^{-\tau(z_{\rm i})}],
\enq
where
\beq
R(z_{\rm i})=\left[\frac{g_un_l(z_{\rm i})}{g_ln_u(z_{\rm i})}-1\right]^{-1}
\left\{\exp\left[\frac{h\nuul}{kT_{\rm r}(z_{\rm i})}\right]-1\right\}.
\enq

For a series of transitions of different frequencies
$\nu^{(k)}$, the total optical depth at any observed frequency $\nu$ is
given by the sum $\tau=\sum_k \tau^{(k)}$ of all the single-line
contributions $\tau^{(k)}$ evaluated at the redshift of interaction
$z_{\rm i}^{(k)}=\nu^{(k)}/\nu$.
 
Remember that for black-body radiation the relative distortion in the
radiation intensity $J_{\nu}$, and in the brightness temperature $T$
are related by
\beq
\frac{\Delta J_\nu}{J_\nu}=\frac{x e^x}{e^x-1}\frac{\Delta T}{T},
\enq
where $x=h\nu/kT$.  
 
\section{radiative and collisional processes}
 
The calculation of the level populations of LiH requires the knowledge
of radiative and collisional rate coefficients  ($R_{ij}$ and $C_{ij}$)
from each state $i$ to any other state $j$. We have included in our
computations only rotovibrational transitions satisfying the selection
rules $\Delta v=0,\pm1$ for $v=0$--10 and $\Delta J=0,\pm 1$ for $J=0$--25.
 
\subsection{Radiative processes}
 
Complete tabulations of the Einstein coefficients for vibro-rotational
transitions of LiH are not available in the literature.  For the sake
of completeness and homogeneity, we have computed all the radiative
transition probabilities by using the appropriate formulae valid for
diatomic molecules in the electric dipole approximation. The rate of
spontaneous emission for pure rotational transitions is given by (e.g.
Shu~1991)
\beq
A_{J\rightarrow J-1}=\frac{64\pi^4\nu^3_{ul}}{3hc^3}D_0^2\frac{J}{2J+1},
\enq
where $D_0=5.888$~debyes (Zemke \& Stwalley~1980) is the value of the
dipole moment evaluated at the equilibrium separation for the two
nuclei. For roto-vibrational transitions,
\beq
A_{(v,J)\rightarrow (v-1,J\pm 1)}=\frac{8\pi^2\nu^3_{ul}}{3h c^3\mu\nu_0}
\left(\frac{{\rm d}D}{{\rm d}r}\right)_0^2 v f_J,
\enq
where
\beq
f_J = \left\{ \begin{array}{ll}
     J/(2J+1)    & \mbox{if $J\rightarrow J-1$,} \nonumber \\
                   (J+1)/(2J+1) & \mbox{if $J\rightarrow J+1$,} \nonumber
              \end{array}
      \right.
\enq
$\nu_0=1405.65$~cm$^{-1}$ (Huber \& Herzberg~1979) is the fundamental
frequency of oscillation of LiH, and $({\rm d}D/{\rm
d}r)_0=0.4132\mbox{e}$ (Zemke \& Stwalley~1980) is the derivative of
the dipole moment evaluated at the equilibrium position.  When a
comparison is possible, the Einstein coefficients computed as indicated
above agree with the accurate but sparse values tabulated by Zemke \&
Stwalley~(1980) to better than 10\%.

The radiative transitions rates $R_{ij}$ can be expressed as function
of the corresponding Einstein coefficients $A_{ij}$ 
\beq R_{ij}=\left\{
\begin{array}{ll} A_{ij}[1+\beta_{ij}(T_{\rm r})]      & \mbox{if
        $i>j$} \\ (g_j/g_i)A_{ji}\beta_{ji}(T_{\rm r}) & \mbox{if 
	$i<j$}
\end{array} 
\right.  
\enq where 
\beq
\beta_{ij}=\left[\exp\left(\frac{E_i-E_j}{kT_{\rm r}}\right)-1\right]^{-1}.
\enq

\subsection{Collisional processes}

We consider only excitation and de-excitation of LiH by collision with
H atoms, and we factorize $C_{ij}=n({\rm H})\gamma_{ij}$.  The cross
sections $\sigma_{0J}$ for the excitation of rotational levels of LiH
by collision with He atoms have been calculated by Jendrek \&
Alexander~(1980) on the basis of the potential surface for the (LiH,
He) system determined by Silver~(1980). These theoretical cross
sections are in good agreement with experimental values (Davies~1986).
The collisional excitation rates $\gamma_{0J}(T_{\rm g})$ can be
obtained from the known cross sections by integrating over a a
maxwellian distribution of particle velocities.  From the cross
sections given by Jendrek \& Alexander~(1980) we have obtained accurate
values of $\gamma_{0J}$ at $T_{\rm g}= 3000$~K.  In order to compute
upward rates for generic $J$ and $J^\prime$ we have made use of the
infinite order sudden factorisation formula (Goldflam et al.~1977) in
the form given by Varshalovich \& Khersonskii~(1977):
\beq
\gamma_{JJ^\prime}(T_{\rm g})=(2J+1)\sum_{\ell=\ell_1}^{\ell_2}\left(
\begin{array}{ccc}
J^\prime & \ell & J \nonumber \\
0        &  0   & 0 \nonumber \\
\end{array}
\right)^2\gamma_{0J}(T_{\rm g}),
\enq
where the summation index $\ell$ assumes integer values between
$\ell_1= |J^\prime-J|$ and $\ell_2=J^\prime+J$, and the term in
parenthesis is a Wigner 3-$J$ symbol.  This expression is valid for
collision energies much larger than rotational energy spacings.
 
The corresponding de-excitation rates are then computed by using the
principle of detailed balance.  The resulting collisional rate
coefficients are of the order of 1--2$\times 10^{-10}$~cm$^3$~s$^{-1}$
at $T_{\rm g}=3000$~K.  Since deexcitation cross sections tend to be
nearly constant at low energies for collisions between neutral
particles (Spitzer~1978), approximate values of $\gamma_{0J}(T_{\rm
g})$ at lower temperatures can be obtained by multiplying the
de-excitation coefficients at $T_{\rm g}=3000$~K by the factor $(T_{\rm
g}/3000~\mbox{K})^{1/2}$.  An additional factor $\sqrt{3}$ must be
adopted to account for the difference in the reduced masses of the
(Li,H) and (Li,He) systems.
%
%
\end{document}